\documentclass[a4paper,10pt,conference]{IEEEtran}
\IEEEoverridecommandlockouts
\usepackage{cite}
\usepackage{amsmath,amssymb,amsfonts}
\usepackage{algorithmic}
\usepackage{graphicx}
\usepackage{textcomp}
\usepackage{xcolor}
\usepackage{comment}
\usepackage{booktabs}
\usepackage{adjustbox}
\def\BibTeX{{\rm B\kern-.05em{\sc i\kern-.025em b}\kern-.08em
    T\kern-.1667em\lower.7ex\hbox{E}\kern-.125emX}}
\begin{document}

\title{
Finding My Way: Influence of Different Audio Augmented Reality Navigation Cues on User Experience and Subjective Usefulness
\thanks{{$\copyright$ \the\year~IEEE. Personal use of this material is permitted. Permission from IEEE must be obtained for all other uses, in any current or future media, including reprinting/republishing this material for advertising or promotional purposes, creating new collective works, for resale or redistribution to servers or lists, or reuse of any copyrighted component of this work in other works. This preprint has not undergone any post-submission improvements or corrections. To cite this article: S. Hinzmann, F. Vona, J. Henning, M. Amer, O. Abdtellatif, T. Kojic and J. -N. Voigt-Antons "Finding My Way: Influence of Different Audio
Augmented Reality Navigation Cues on User
Experience and Subjective Usefulness," 2025 IEEE International Conference on Quality of Multimedia Experience, Madrid, Spain, 2025, pp. to appear, doi: to appear}}
}

%
%

\author{
\IEEEauthorblockN{Sina Hinzmann$^{1}$, Francesco Vona$^{1}$, Juliane Henning$^{1}$, Mohamed Amer$^{1}$, Omar Abdellatif$^{1}$, Tanja Kojic$^{2}$, Jan-Niklas Voigt-Antons$^{1}$}
\IEEEauthorblockA{$^1$Immersive Reality Lab, Hamm-Lippstadt University of Applied Sciences, Hamm, Germany\\
$^2$Quality and Usability Lab, Technische Universität Berlin, Berlin,  Germany
}
\author{
\IEEEauthorblockN{Sina Hinzmann$^{1}$, Francesco Vona$^{1}$,  Juliane Henning$^{1}$\\
Mohamed Amer$^{2}$, 
Omar Abdellatif$^{2}$, 
Tanja Kojic$^{3}$, 
Jan-Niklas Voigt-Antons$^{1,3}$\\}
\IEEEauthorblockA{\textit{$^{1,2}$Hamm-Lippstadt University of Applied Sciences, Hamm, Germany}
\textit{$^3$Technische Universität Berlin, Berlin, Germany}\\
$^1$\{firstname.lastname\}@hshl.de,
$^2$ \{firstname.lastname\}@stud.hshl.de,
$^3$\{firstname.lastname\}@tu-berlin.de}}}

\maketitle
\begin{abstract}
As augmented reality (AR) becomes increasingly prevalent in mobile and context-aware applications, the role of auditory cues in guiding users through physical environments is becoming critical. This study investigates the effectiveness and user experience of various categories of audio cues, including fully non-verbal sounds and speech-derived Spearcons, during outdoor navigation tasks using the Meta Quest 3 headset. Twenty participants navigated five outdoor routes using audio-only cue types: Artificial Sounds, Nature Sounds, Spearcons, Musical Instruments, and Auditory Icons. Subjective evaluations were collected to assess the perceived effectiveness and user experience of each sound type. Results revealed significant differences in perceived novelty and stimulation across sound types. Artificial Sounds and Musical Instruments were rated higher than Spearcons in novelty, while Artificial Sounds were also rated higher than Spearcons in stimulation. Overall preference was evenly split between Nature Sounds and Artificial Sounds. These findings suggest that incorporating aspects of novelty and user engagement in auditory feedback design may enhance the effectiveness of AR navigation systems.
\end{abstract}


\section{Introduction}
The integration of auditory interfaces in augmented reality (AR) applications has become increasingly relevant as immersive technologies migrate from controlled lab settings to real-world, mobile contexts \cite{b7,b17,b18,b19}. Auditory navigation systems offer a promising alternative where visual cues are impractical, distracting, or cognitively demanding \cite{b9,b21}. This demand is evident in scenarios with environmental constraints, accessibility requirements, or multitasking needs, where users must navigate complex spaces using only audio~\cite{b9,b19, b20,b24}.
Conventional navigation aids primarily rely on speech-based instructions—such as those found in GPS smartphone applications—which can lead to cognitive overload, language barriers, and increased attentional demands~\cite{b5,b11,b22}. In contrast, brief non-verbal auditory cues—including artificial tones, natural sounds, musical signals, environmental icons, and spearcons (time-compressed speech)~\cite{b1,b3,b10,b12,b13}—may convey navigation information more efficiently through learned or intuitive associations.
Although previous research has explored individual verbal and non-verbal cues in navigation and HCI, few studies have systematically compared diverse audio cue categories within realistic outdoor AR settings~\cite{b6,b7,b21}. Furthermore, user perceptions of these cues remain underexplored during actual navigation tasks~\cite{b19,b20}. To address these gaps, the present study evaluates the effectiveness and perceived quality of five distinct categories of audio cues for navigation in an audio-only AR application developed for the Meta Quest 3 headset.
A pre-study was conducted to identify user-preferred sounds for typical navigation scenarios. The resulting cue set was employed in the main experiment, where participants completed a series of audio-guided navigation tasks under each cue condition. Both performance data and user experience ratings were collected to assess the utility and user acceptance of each cue type.

\section{Related Work}
Auditory displays have long been explored in HCI and cognitive ergonomics as an effective means of conveying information without overloading the visual channel. Foundational categories of structured auditory displays, including earcons, a brief, non-speech audio message that conveys information and auditory icons, established early design principles for symbolic audio interaction~\cite{b1}. This work was expanded by the introduction of ecological auditory cues, emphasizing the intuitive mapping of familiar environmental sounds to interface events~\cite{b2, b12}.
The development of spearcons, ultra-short, time-compressed speech-derived cues, provided a hybrid solution for rapid auditory recognition without the need for full verbal comprehension~\cite{b3, b10}. Comprehensive reviews have demonstrated how different auditory cues impact cognitive load, information processing time, and user trust, especially in mobile and multitasking contexts~\cite{b4, b9}. Laboratory investigations have further shown that spearcons can outperform earcons in recognition speed in certain applications, such as patient monitoring~\cite{b10}.
With the increasing adoption of spatial computing and AR, the use of audio cues for navigation has received growing attention. Auditory icons have been shown to support non-visual navigation in both desktop and real-world environments~\cite{b5, b6}. Research on spatial audio AR has addressed orientation measurement challenges and demonstrated improvements in navigational awareness and cognitive load reduction~\cite{b7, b17, b18}. The design of audio-based and non-visual interfaces for AR continues to advance, with recent reviews highlighting the influence of audio cue design on user experience and accessibility~\cite{b19, b20,b22,b24}. Further research examines environmental acoustics and playback devices affecting cue effectiveness in real-world settings~\cite{b21}, while reviews summarize technologies, applications, and outcomes in this domain~\cite{b20,b22}.
Recent work has also demonstrated that spatial auditory navigation can have a significant impact on user experience during augmented outdoor navigation tasks~\cite{b7}. However, there remains a lack of direct comparisons between distinct categories of audio cues in real-world, eyes-free AR navigation.

\section{Methods}
\subsection{Experimental Design}
We employed a within-subjects experimental design where each participant completed five audio-only navigation tasks, each guided by a different sound cue category. The order of conditions was randomized using a Latin Square design to control for order effects. For each condition, participants were guided through a unique navigation route, ensuring that no path was repeated and that route familiarity did not bias the results. A subjective evaluation was obtained through post-task questionnaires, which included the User Experience Questionnaire (UEQ) \cite{b15,b16} as well as custom questions developed for the present experiment. Participants responded to the statements: “I had a positive overall navigation experience with the audio cues (Q1),” “I was able to clearly perceive the direction of the audio cues (Q2),” “I understood the meaning of the audio cues (Q3),” “The audio cues helped me navigate effectively (Q4),” and “I found the audio cues pleasant to use (Q5).” If a task exceeded six minutes, the condition was marked as failed. In addition, three open-ended questions were included to elicit qualitative feedback: “What did you like about this type of audio cue?”, “What did you not like about this type of audio cue?”, and “How could this type of audio cue be improved for more efficient or pleasant navigation?”. After all conditions, participants were asked to indicate their preferred sound condition by selecting one of the five sound cues. The study received ethical approval from the local ethics committee.
\subsection{Pre-Study: Identifying Preferred Navigation Sounds}
Before the main study, we conducted a pre-study with 31 participants to identify the most intuitive and user-preferred sounds for each cue category. Participants were seated in a quiet lab environment and wore over-ear headphones connected to a desktop interface. They were presented with four prototypical navigation scenarios: i) “Turn left at the next intersection”, ii) “Obstacle ahead – change direction”, iii) “Continue straight for 100 meters”, iv) “You have reached your destination”.
For each scenario, participants listened to 15 audio cues (3 from each of the 5 categories: Artificial Sounds, Nature Sounds, Spearcons, Musical Instruments, and Auditory Icons), presented in randomized order. These categories were selected because they reflect the wide diversity in prior auditory interface research. After each group of three within a category, participants were asked to vote for the sound they believed was most effective and intuitive for the given instruction. The highest-rated sounds in each category formed the final stimulus set for the main study.
\subsection{Participants}
A total of 20 participants were initially recruited for the study. Five participants were excluded from the analysis for not completing at least two experimental conditions to allow for meaningful within-subject comparisons. The final sample thus consisted of 15 participants (3 female, 12 male), with ages ranging from 20 to 45 years (\textit{M} = 28.33, \textit{SD} = 8.37). 
Participants reported their prior experience with virtual, augmented, and mixed reality technologies: none (\textit{n} = 4), little (\textit{n} = 6), some (\textit{n} = 2), and extensive (\textit{n} = 3). The mean Affinity for Technology Interaction (ATI) score was \textit{M} = 4.27 (\textit{SD} = 0.92, $\alpha = 0.85$), indicating a moderate to moderately high level of technology affinity and a good internal consistency.
\begin{table*}[!ht]
\centering
\caption{
Mean and standard deviation for each sound condition on the six UEQ subscales and five custom questionnaire items. 
}
\label{tab:results_table}

\resizebox{1\textwidth}{!}{
\begin{tabular}{l*{11}{c}}
\toprule
& \multicolumn{6}{c}{\textbf{UEQ}} & \multicolumn{5}{c}{\textbf{Custom Questions}} \\
\cmidrule(lr){2-7} \cmidrule(lr){8-12}
\textbf{Condition} & \textbf{Attractiveness} & \textbf{Perspicuity} & \textbf{Efficiency} & \textbf{Novelty} & \textbf{Stimulation} & \textbf{Dependability} & \textbf{Q1} & \textbf{Q2} & \textbf{Q3} & \textbf{Q4} & \textbf{Q5} \\
\midrule
Artificial Sounds  & 1.26 (1.17)   & 1.05 (1.44) & 1.20 (1.07)  & 0.85 (0.80)  & 1.17 (1.23) & 0.76 (1.29)    & 5.13 (1.85) & 5.00 (1.81) & 5.60 (1.45) & 5.21 (1.63) & 5.27 (1.22) \\
Auditory Icons & 0.72 (1.11)   & 0.89 (1.44)  & 0.83 (1.05) & 0.53 (0.92)  & 0.58 (1.02) & 0.58 (0.94)    & 4.86 (1.75) & 5.14 (1.96) & 5.21 (1.76) & 5.29 (1.33) & 4.43 (1.60)  \\
Music Instruments  & 1.19 (1.13)   & 0.83 (1.39)  & 0.78 (1.21) & 0.98 (0.65)  & 1.05 (1.03) & 0.76 (1.22)    & 5.20 (1.70) & 5.20 (1.93) & 4.93 (1.75) & 5.53 (1.36) & 5.27 (1.62)  \\
Nature Sounds & 0.81 (1.27)   & 0.41 (1.67) & 0.61 (1.38) & 0.96 (1.10)  & 0.86 (1.00) & 0.41 (1.14)    & 4.47 (1.92) & 4.60 (1.76) & 4.33 (1.84) & 5.00 (1.65) & 5.20 (1.74)  \\
Spearcons  & 0.48 (1.28)   & 1.33 (1.13)  & 1.08 (1.25) & -0.36 (1.06) & 0.26 (1.48) & 0.98 (1.36)    & 4.87 (1.55) & 5.00 (1.46) & 6.07 (1.22) & 5.40 (1.59) & 4.33 (1.54) \\
\bottomrule
\end{tabular}%
}
\end{table*}

\subsection{Setup and Procedure}

The study took place in a controlled outdoor area on the university campus with low pedestrian and no vehicle traffic, ensuring participant safety. The setting provided realistic conditions for auditory navigation, including typical ambient noise and natural acoustics. To ensure focus on auditory perception and prevent potential motion sickness, participants wore the Meta Quest 3 headset on their forehead for head tracking only, and used open-back high-fidelity headphones for spatialized audio cues delivered by a custom Unity AR application. Each participant navigated five checkpoints per route (totaling 55 meters), with route orientation controlled and no repetition within a session. The full session lasted approximately 60 minutes and included: i) onboarding, consent, and demographics; ii) navigation tasks with post-task questionnaires (repeated for each audio cue category); and iii) debriefing.

\section{Results} 
\begin{figure}[!ht]
\centerline{\includegraphics[width=0.45\textwidth]{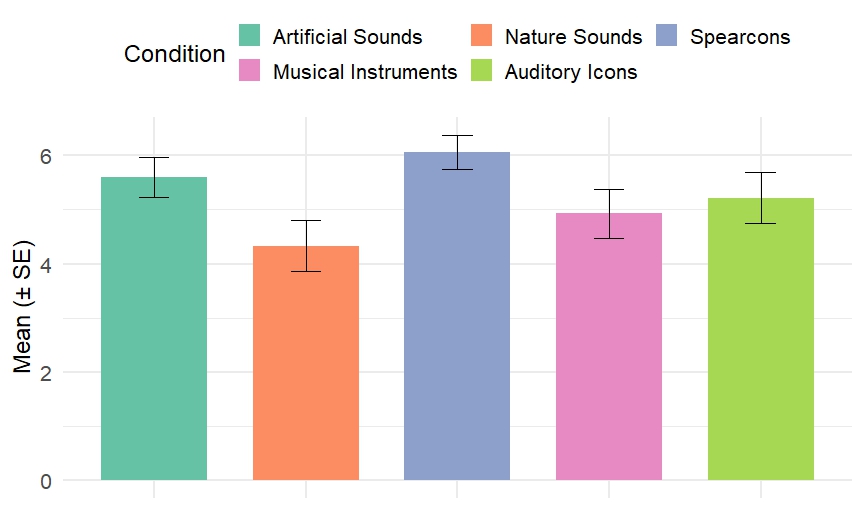}}
\caption{Mean responses (±SE) to custom question Q3 (“I understood the meaning of the audio cues”) for each sound condition. 
}
\label{fig:results2}
\end{figure}

Due to the relatively small sample size, all analyses employed non-parametric statistics to ensure robust inference. Only statistically significant findings are reported below.
\begin{figure}[!ht]
\centerline{\includegraphics[width=0.45\textwidth]{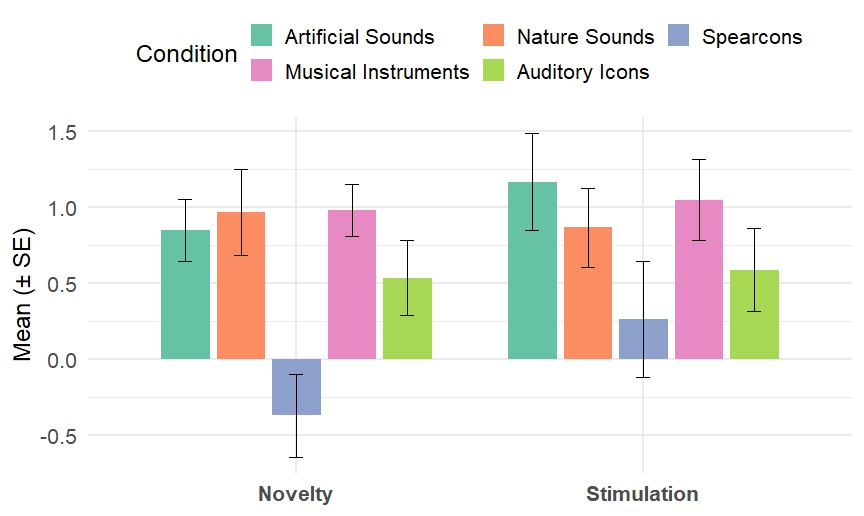}}
\caption{Mean user ratings (±SE) for the UEQ subscales \textit{Novelty} and \textit{Stimulation} across the five sound conditions. 
}
\label{fig:results1}
\end{figure} 
User experience ratings were compared across the five sound conditions—Artificial Sounds, Nature Sounds, Spearcons, Musical Instruments, and Auditory Icons—on the six UEQ scales: \textit{Attractiveness}, \textit{Perspicuity}, \textit{Efficiency}, \textit{Novelty}, \textit{Stimulation}, and \textit{Dependability}. Friedman tests revealed significant differences for the scales \textit{Novelty} ($\chi^2$(4) = 14.20, $p$ = .007) and \textit{Stimulation} ($\chi^2$(4) = 9.52, $p$ = .049). Post hoc Wilcoxon signed-rank tests with Holm correction showed that both Artificial Sounds and Musical Instruments were rated significantly higher than Spearcons on \textit{Novelty} ($p_\mathrm{adj}$ = .023 for both), while Artificial Sounds also scored significantly higher than Spearcons on \textit{Stimulation} ($p_\mathrm{adj}$ = .031; see Fig.~\ref{fig:results1}). Although the pairwise comparison between Nature Sounds and Spearcons on the \textit{Novelty} scale was statistically significant before correction ($p$ = .006), the result did not remain significant after Holm correction ($p_\mathrm{adj}$ = .051). 
A Friedman test identified a significant effect for Q3 ($\chi^2$(4) = 9.90, $p$ = .042; see Fig.~\ref{fig:results2}), though subsequent pairwise Wilcoxon signed-rank tests with Holm correction revealed no significant differences between conditions. 
Regarding overall preference, participants’ final votes showed a tie between Nature Sounds (\textit{n} = 5) and Artificial Sounds (\textit{n} = 5), with Musical Instruments the next most preferred (\textit{n} = 3), followed by Auditory Icons (\textit{n} = 1). Spearcons received no votes.


\section{Discussion and Conclusion} 

This study systematically compared user experience across five categories of auditory cues for AR navigation: Artificial Sounds, Nature Sounds, Spearcons, Musical Instruments, and Auditory Icons. 

Results from the UEQ revealed significant differences in perceived novelty and stimulation among the cue types. Specifically, Spearcons were rated significantly lower in both novelty and stimulation compared to Artificial Sounds, and lower in novelty compared to Musical Instruments. This likely reflects the familiarity of spearcons, which resemble conventional speech-based navigation prompts, reducing their perceived originality~\cite{b3, b10}. In contrast, Artificial Sounds and Musical Instruments—less commonly used in navigation contexts—were seen as more novel and engaging. The abstract and unfamiliar qualities of Artificial Sounds, in particular, may have contributed to higher stimulation ratings, supporting prior findings that novelty enhances user engagement with auditory interfaces~\cite{b19}.
Despite these perceptual differences, subjective effectiveness and objective task completion times did not vary significantly between cue categories, indicating similar functional performance. Future research could include additional measures such as navigation accuracy or error rates to further assess cue effectiveness.
Regarding overall preference, participants most frequently chose Nature Sounds and Artificial Sounds, echoing qualitative feedback that described these categories as “pleasant” and “relaxing.” The beneficial effects of nature sounds are well established~\cite{b23}, and a near-significant difference in novelty ratings was observed between Nature Sounds and Spearcons. Musical Instruments were also positively received, often described as friendly and pleasant.
Notably, no participant selected Spearcons as their preferred category, diverging from previous findings in menu-based auditory interfaces where spearcons have been shown to improve navigation~\cite{b3, b10}. This difference may be due to the challenging listening conditions of outdoor navigation, where environmental noise can diminish the clarity and impact of time-compressed speech cues. Participant feedback indicated a preference for slower, more natural-sounding cues, suggesting that the pace and familiarity of spearcons limited their appeal in this setting. However, the relatively small sample size (\textit{n} = 15) may have constrained the detection of more subtle effects. While the within-subject design and use of non-parametric tests helped address some limitations, larger studies are needed to confirm and extend these findings.
Based on these results, we recommend that future spatial audio navigation systems employ continuous, clearly distinguishable auditory cues that are easily audible and free from unnecessary pauses or ambiguities. Our findings also highlight users’ openness to novel and unconventional sound types for navigation, suggesting opportunities for further innovation in auditory interface design.

\end{document}